\documentclass[10pt]{wlscirep}
\usepackage[resetlabels]{multibib}
\usepackage{bm}
\usepackage{lineno}
\usepackage{mathpazo}
\usepackage{amsfonts,amssymb,stmaryrd,latexsym,amsmath,braket}
\usepackage{graphicx,subfigure}
\usepackage[export]{adjustbox}
\usepackage{times}
\usepackage{bbm}
\usepackage{bm}
\usepackage{mathrsfs}
\usepackage{slashed}
\usepackage[version=3]{mhchem}
\usepackage{braket}
\usepackage{hyperref}
\usepackage{verbatim}
\usepackage[utf8]{inputenc}

\usepackage{graphicx,subfigure}
\usepackage{times}
\usepackage{slashed}
\usepackage{braket}
\usepackage{verbatim}
\usepackage{multirow}
\usepackage[utf8]{inputenc}
\usepackage{array}
\usepackage{tensor}
\newcommand{\PreserveBackslash}[1]{\let\temp=\\#1\let\\=\temp}
\newcolumntype{C}[1]{>{\PreserveBackslash\centering}p{#1}}
\newcolumntype{R}[1]{>{\PreserveBackslash\raggedleft}p{#1}}
\newcolumntype{L}[1]{>{\PreserveBackslash\raggedright}p{#1}}

\makeatletter
\def\namedlabel#1#2{\begingroup
   \def\@currentlabel{#2}%
   \label{#1}\endgroup
}
\makeatother

\title{Machine learning orbital-free density functional theory: taming quantum shell effects in deformed nuclei}

\author[1,2]{X. H. Wu}
\author[2,3,4]{Z. X. Ren}
\author[2]{P. W. Zhao\footnote{pwzhao@pku.edu.cn}}

\affil[1]{Department of Physics, Fuzhou University, Fuzhou 350108, Fujian, China}

\affil[2]{State Key Laboratory of Nuclear Physics and Technology, School of Physics, Peking University, Beijing 100871, China}

\affil[3]{Institute for Advanced Simulation, Forschungszentrum J\"{u}lich, D-52425 J\"{u}lich, Germany}

\affil[4]{Helmholtz-Institut f\"{u}r Strahlen- und Kernphysik and Bethe Center for Theoretical Physics, Universit\"{a}t Bonn, D-53115 Bonn, Germany}


\begin{abstract}
Accurate description of deformed atomic nuclei by the orbital-free density functional theory has been a longstanding textbook challenge, due to the difficulty in accounting for the intricate quantum shell effects that are present in such systems.
Orbital-free density functional theory is, in principle, capable of describing all effects of nuclear systems, as guaranteed by the Hohenberg-Kohn theorem. 
However, from a microscopic perspective, shell and deformation effects are believed to be intrinsically connected to single-orbital structures, posing a significant challenge for orbital-free approaches.
Here,  we develop a machine learning approach to the orbital-free density functional theory, which is capable of achieving a high level of accuracy in describing  the ground-state properties and potential energy curves for both spherical $^{16}$O and deformed $^{20}$Ne nuclei.
This is the inaugural instance where a fully orbital-free energy density functional has succeeded in taming the complex shell effects in deformed nuclei. It demonstrates that the orbital-free energy density functional, which is directly based on the Hohenberg-Kohn theorem, is not only a theoretical concept but also a practical one for nuclear systems.

\end{abstract}

\begin{document}


\maketitle

Shell effects are general characteristics for finite quantum many-body systems, including atoms, nuclei, confined quantum gases and fluids, nanostructures, quantum dots, and other similar entities.
Quantum shell effects are typically associated with a significant energy gap in the single-particle energy spectrum near the Fermi level. Such gaps provide additional binding energies and enhance the stability of the systems.
In nuclear systems, the shell effects are intimately connected to nuclear deformation, which arises from the spontaneous symmetry breaking of the nuclear mean field in the intrinsic system~\cite{Nilsson1955}.
Therefore, a precise and self-consistent description of nuclear shell and deformation effects is a pivotal aspect of nuclear theory.

In 1964, Hohenberg and Kohn demonstrated that the total energy of a multibody system can be expressed as a functional of density~\cite{Hohenberg1964Phys.Rev.}.
Consequently, the quantum many-body problem formulated in terms of $N$-body wave functions could be mapped into a one-body level with the density distribution.
This led to the development of Density Functional Theory (DFT), which has become a highly active area in various fields, including quantum chemistry, condensed-matter physics, and nuclear physics.
The DFT is in principle capable of describing all effects of nuclear systems with the guarantee of the Hohenberg-Kohn theorem.
However, from a microscopic perspective, the shell and deformation effects are related to the bifurcations of single-particle levels and, thus, they are always incorporated into the energy density functional by the introduction of auxiliary one-body orbits, i.e., the Kohn-Sham DFT~\cite{Kohn1965Phys.Rev.}.
It remains unclear how to incorporate shell and deformation effects into the orbital-free energy density functional directly based on the Hohenberg-Kohn theorem without the introduction of the auxiliary one-body orbits in practice.

The current attempts at orbital-free DFT are mainly based on the semiclassical Thomas-Fermi (TF) approach and its extended versions (ETF)~\cite{Brack1985Phys.Rep., Centelles2007Ann.Phys.}. 
However, in these semiclassical approaches, all nuclei are spherical in their ground states~\cite{Brack1985Phys.Rep., Colo2023PTEP}, and the obtained ground-state densities lack the quantum shell and deformation effects~\cite{Brack1985Phys.Rep., Centelles1990Nucl.Phys.A, Centelles2007Ann.Phys.}.
To incorporate the shell and deformation effects, additional corrections must be applied~\cite{Dutta1986Nucl.Phys.A, Aboussir1995Atom.DataNucl.DataTables}, i.e., the shell corrections~\cite{Brack1972Rev.Mod.Phys., Brack1973Nucl.Phys.A}.
Two well-known methods for incorporating shell effects are the Strutinsky-integral method~\cite{Strutinsky1967NPA, Strutinsky1968NPA, Brack1975PLB} and the expectation-value method~\cite{Bohigas1976Phys.Lett.B, Chu1977Phys.Lett.B}. 
However, both methods involve the auxiliary one-body orbits, which means that the ETF approaches with additional shell corrections are no longer orbital-free DFTs that are directly based on the Hohenberg-Kohn theorem. An accurate description of the shell and deformation effects by the orbital-free DFT has remained an elusive topic for nuclear physics. In principle, this is guaranteed by the Hohenberg-Kohn theorem. However, in practice, this has not yet been achieved, and moreover, the feasible route to achieving this remains unclear for a long time.

Machine learning (ML) has been successfully applied in various fields of physics~\cite{Carleo2019Rev.Mod.Phys., Boehnlein2022Rev.Mod.Phys., He2023Sci.ChinaPhys.Mech.Astron.}.
In nuclear physics, ML has been widely adopted to nuclear masses~\cite{Niu2013Phys.Rev.C, Utama2016Phys.Rev.C, Niu2018Phys.Lett.B, Neufcourt2018Phys.Rev.C, Pastore2020Phys.Rev.C, Wu2020Phys.Rev.C051301, Wu2021Phys.Lett.B, Wu2022Phys.Lett.B137394, Niu2022Phys.Rev.C},
charge radii~\cite{Akkoyun2013J.Phys.GNucl.Part.Phys., Utama2016J.Phys.GNucl.Part.Phys., Ma2020PRC, WuDi2020Phys.Rev.C, Ma2022Chin.Phys.C,Dong2023PLB},
decays and reactions~\cite{Niu2019Phys.Rev.Ca, Lovell2020J.Phys.GNucl.Part.Phys., WuDi2021Phys.Rev.C, Saxena2021J.Phys.GNucl.Part.Phys., Neudecker2021Phys.Rev.C, Wang2021Chin.Phys.Ca, Huang2022Commun.Theor.Phys.},
ground and excited states~\cite{Jiang2019Phys.Rev.C, Lasseri2020Phys.Rev.Lett., Yoshida2020Phys.Rev.C, Wang2021Phys.Rev.Ca, Bai2021Phys.Lett.B},
nuclear landscape~\cite{Neufcourt2019Phys.Rev.Lett., Neufcourt2020Phys.Rev.Ca},
fission yields~\cite{Wang2019Phys.Rev.Lett., Lovell2020J.Phys.GNucl.Part.Phys.a, Qiao2021Phys.Rev.C},
nuclear many-body calculations~\cite{Keeble2020Phys.Lett.B, Adams2021Phys.Rev.Lett., Lovato2022PRR, Yang2022PLB, Yang2023PRC, Rigo2023PRE}, etc.
ML is a powerful tool for finding existing and complicated patterns in high-dimensional data, which is very suitable for constructing the energy density functional (EDF), i.e., a functional that is proven to exist but of unknown form.
There has been considerable works on the application of machine learning in constructing orbital-free DFT for electronic systems~\cite{Pederson2022NRP, Huang2023Science}.
A proof-of-principle for ML-DFT was demonstrated in 2012~\cite{Snyder2012Phys.Rev.Lett.}, showing that ML methods can achieve an excellent approximation of the kinetic energy density functional for non-interacting fermions in a 1D box.
The practical usefulness of ML in DFT was later demonstrated through realistic examples~\cite{Brockherde2017Nat.Commun.}. 
Since then, numerous efforts have been made to bring the promise of ML to practical, generalizable functional constructions.
In Ref.~\cite{Nagai2020npjCM}, it was shown that a neural network functional trained on accurate densities and energies of just three molecules can perform as well as human-designed functionals for 150 test molecules, exhibiting excellent generalization ability.
It was demonstrated that density-based $\Delta$-learning (learning only the correction to a standard DFT calculation) can significantly reduce the amount of training data required~\cite{Bogojeski2020Nat.Commun.}.
The learning of Hohenberg-Kohn maps in DFT was found to be less effective across quantum phase transitions, suggesting an intrinsic challenge in efficiently learning non-smooth functional relationships~\cite{Moreno2020Phys.Rev.Lett.}.
It was also found that incorporating prior knowledge during training can enhance the generalization capability of the functional~\cite{Li2021Phys.Rev.Lett.}.
A DeepMind team provided the DM21 functional, which was trained on thousands of molecular systems and outperforms most other hybrid functionals on standard molecular benchmarks, with impressive generalization~\cite{Kirkpatrick2021Science}.
A density-fitting representation was introduced in ML-based DFT instead of the real-space grid representation~\cite{Margraf2021Nat.Commun.}, and was applied to real molecules.
A machine representation of symbolic functionals was proposed~\cite{Ma2022SA}, which is more interpretable to humans. 
Machine-learning functionals for excited-state dynamics simulations were also developed~\cite{Bai2022NC}.

For nuclear systems, the ML application of orbital-free DFT is still in its infancy~\cite{Wu2022Phys.Rev.C, Hizawa2023PRC}.
Recently, we have successfully employed the ML to construct a robust and accurate orbital-free EDF for spherical nuclei.
Self-consistent calculations with this ML orbital-free EDF bypassing the Kohn-Sham equations provide the ground-state densities, total energies, and root-mean-square radii of $^{4}$He, $^{16}$O and $^{40}$Ca, with a high accuracy in reproducing the orbital-dependent Kohn-Sham solutions~\cite{Wu2022Phys.Rev.C}.
This proves, in practice, the feasibility of constructing highly accurate nuclear orbital-free EDFs via the ML approach.
Therefore, one could expect that the ML approach can also help to construct the missing shell and deformation effects in the orbital-free DFT.

In this work, we directly address the challenge of constructing orbital-free EDFs to describe the shell effects in deformed atomic nuclei.
The ML kernel ridge regression (KRR) approach is adopted and the model is trained to build the map from the nucleon density onto both the kinetic and spin-orbit energies.
Together with the interaction energy taken from the Skyrme functional, an orbital-free EDF for deformed nuclei is constructed.
The ML orbital-free EDF is then used to calculate the ground-state properties of the spherical $^{16}$O and deformed $^{20}$Ne nuclei, and the results are compared with the Kohn-Sham calculations and the experimental data.
$^{20}$Ne is a prototypical deformed nucleus, customarily used as an example to illustrate the transition from spherical to deformed nuclear systems in newly developed approaches~\cite{Vautherin1973Phys.Rev.C, Lee1986PhysRevLett.57.2916, Pannert1987PhysRevLett.59.2420}.
Moreover, constrained calculations are realized in the framework of orbital-free DFT, and the potential energy curves are obtained for both $^{16}$O and $^{20}$Ne as functions of the quadrupole deformation.

In the framework of the orbital-free DFT, the total energy of a nuclear system can be expressed as a functional of the density alone,
\begin{equation}\label{Etot1}
  E_{\rm tot}[\rho] = E_{\rm kin}[\rho] + E_{\rm int}[\rho].
\end{equation}
The interaction energy here is taken from the Skyrme functional SkP~\cite{Dobaczewski1984Nucl.Phys.A103}, i.e., $E_{\rm int}[\rho] = \int \mathcal{E}_{\rm Skyrme} {\rm d}{\bm r}$, where
\begin{align}
  \mathcal{E}_{\rm Skyrme} = & \frac{3}{8}t_0\rho^2 + \frac{1}{16}t_3\rho^{2+\gamma} + \frac{1}{64}(9t_1 - 5t_2 - 4t_2x_2)(\nabla\rho)^2 \nonumber \\
  & -\frac{1}{16}\left[ (t_1x_1 + t_2x_2) - \frac{1}{2} \right]{\bm J}^2 + \frac{3}{4}W_0{\bm J}\nabla\rho. \label{Eint}
\end{align}
The Skyrme functional contains two parts, i.e., the interaction part $E_{\rm int}'$ (the first line of Eq.~\eqref{Eint}), which depends explicitly only on the nucleon density $\rho$, and the spin-orbit part $E_{\rm so}$ (the second line of Eq.~\eqref{Eint}), which depends additionally on the spin-orbit density ${\bm J}$.
To build an orbital-free EDF, the spin-orbit and kinetic energies are expressed in terms of the nucleon density within the KRR method,
\begin{equation}\label{KRR_kinso}
  E_{\rm kin+so}^{\rm ML}[\rho] = \sum_{i=1}^{m} \omega_{i} K(\rho_i,\rho).
\end{equation}
Here, $\rho_i({\bm r})$ are training densities, $K$ is the kernel function which measures the similarity between densities, and $\omega_i$ are weights to be determined in the KRR framework by
\begin{equation}\label{weight_paras}
  {\bm \omega} = ({\bm K}+\lambda{\bm I})^{-1}{\bm E}_{\rm kin+so},
\end{equation}
where $\lambda$ is a regularizer that penalizes large weights to reduce the risk of overfitting, ${\bm I}$ is the identity matrix, and ${\bm E}_{\rm kin+so}$ are the exact kinetic and spin-orbit energies to be learned, i.e., $(E_{\rm kin+so}[\rho_1], \ldots, E_{\rm kin+so}[\rho_m])$, ${\bm K}$ is the kernel matrix with elements ${\bm K}_{ij} = K(\rho_i,\rho_j)$,
\begin{equation}\label{Kernel}
  K(\rho,\rho') = \exp\left[ -||\rho(\bm{r})-\rho'(\bm{r})||^2/(2A A'\sigma^2) \right].
\end{equation}
Here, $\sigma$ is a hyperparameter defining the length scale on the distance that the kernel affects, and the distance between two densities $||\rho(\bm{r})-\rho'(\bm{r})||$ can be calculated by vectorizing the densities on a series of discrete grids. The factors $A$ and $A'$ are
the nucleon numbers of the densities $\rho$ and $\rho'$, respectively.

For a given set of the training data $(\rho, E_{\rm kin}, E_{\rm so})$, the weight parameters \eqref{weight_paras} can be determined, and then the KRR predictions~\eqref{KRR_kinso} of the sum of the kinetic and spin-orbit energies are obtained. Taking the interaction energy into account, the orbital-free EDF is written as
\begin{equation}\label{Etot2}
  E_{\rm tot}^{\rm ML}[\rho] = E_{\rm kin+so}^{\rm ML}[\rho] + E_{\rm int}'[\rho],
\end{equation}
where $E_{\rm int}'[\rho]$ denotes the interaction energy without the spin-orbit part.
The nuclear ground state is then obtained by minimizing the energy density functional \eqref{Etot2} with respect to the density.

In this work, we consider axially deformed nuclei and, thus, the nucleon density can be written in two dimension with the metric $\tilde{\rho}(r_{\bot},z) = 2\pi r_{\bot}\rho(r_{\bot},z)$. Numerically, the nucleon density is expressed in 1128 discretized spatial mesh points of $(r_{\bot},z)$ with $r_{\bot}\in[0,\,8.05]$~fm and $z\in[-8.05,\, 8.05]$~fm. The training data of densities, kinetic energies, and spin-orbit energies are prepared by solving the Schr\"odinger equations with randomly generated mean potentials including the spin-orbit potential,
\begin{equation}\label{Sch_eq}
  \left[ -\frac{\hbar^2}{2m} + V + {\bm W}\cdot(-i)(\nabla\times\sigma) \right]\psi = E\psi.
\end{equation}
The mean potential $V$ is simulated by a combination of spherical $V_0({\bm r})$ and quadrupole $V_2({\bm r})$ components.
The radial functions of these two components are simulated by combined Gaussian functions with 6 parameters~\cite{Wu2022Phys.Rev.C}, which are randomly generated in a proper range.
The spin-orbit potential is given by ${\bm W}({\bm r}) = \frac{3}{4}W_0{\bm \nabla}\rho_{\rm pre}$, where the density $\rho_{\rm pre}$ is pre-calculated by solving the Schr\"odinger equation with ${\bm W}({\bm r}) = 0$.

Given that the Schr\"odinger equation can be easily solved, it is possible to generate a large database with minimal effort.
In this work, we generate 24,000 data for the nucleon densities, kinetic and spin-orbit energies of spherical $^{16}$O and deformed $^{20}$Ne.
The data are divided into three sets, comprising 20,000 data for the training set, 2,000 for the validation set, and 2,000 for the test set.
In each set, the numbers of the data for systems with $A=16$ and $A=20$ are identical.

The KRR network~(\ref{KRR_kinso}) is trained with the training data, and the solution can be obtained via Eq.~(\ref{weight_paras}).
Here, the hyperparameters $\lambda$ and $\sigma$ are determined by optimizing the ML performance on the validation set.
The resulting hyperparameters are $\lambda = 2.19\times 10^{-6} $ and $\sigma = 1.86$ fm$^{-2}$.
Finally, the test set of 2000 data samples is used to provide an unbiased evaluation of the KRR training.

\begin{figure}[!htbp]
\centering
\includegraphics[width=8.0cm]{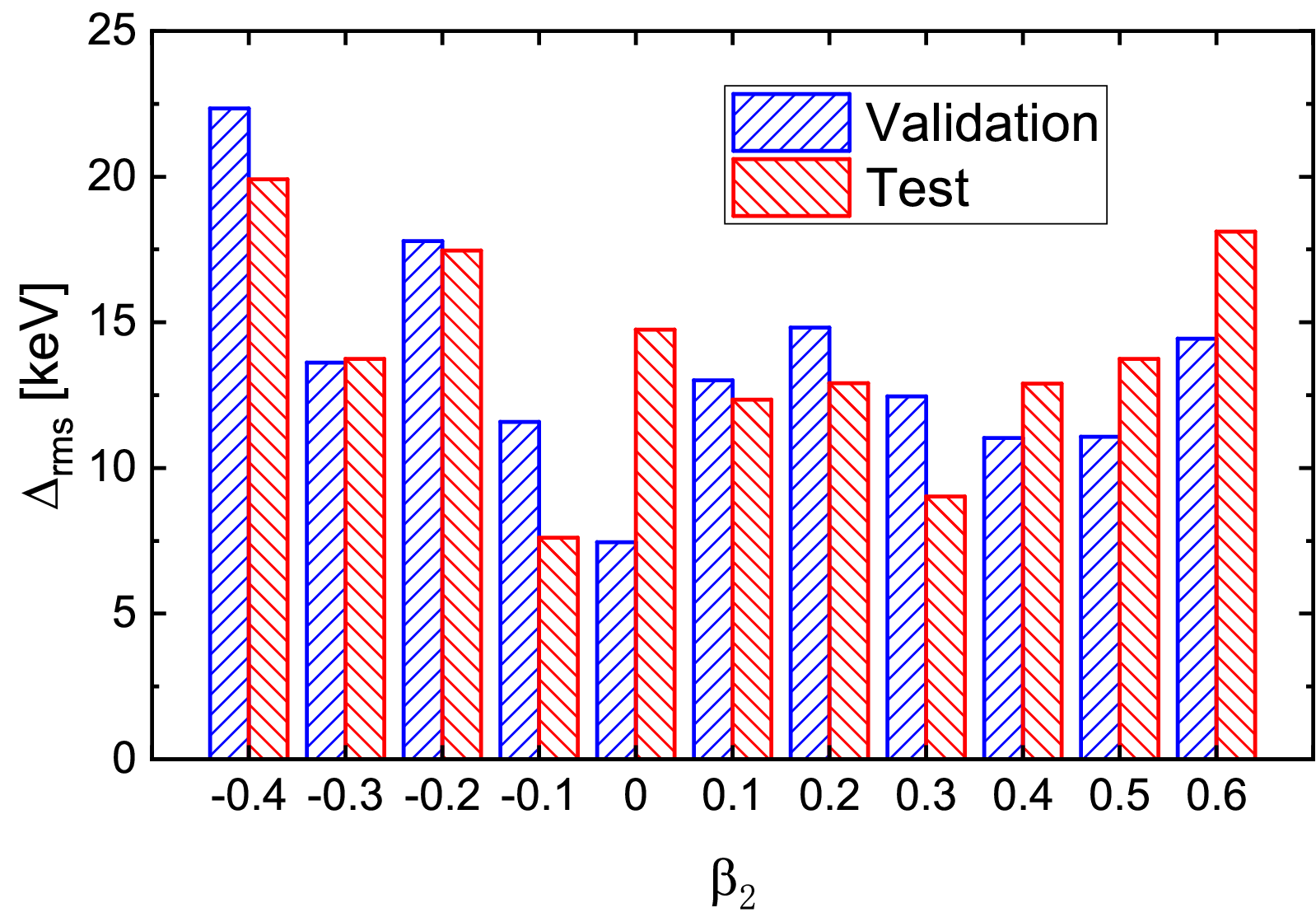}
\caption{The root-mean-square (rms) deviations $\Delta_{\rm rms}$ of the kinetic and spin-orbit energies between the KRR predicted results and the exact values in the validation and test sets.
The quadrupole deformations $\beta_2$ are in different slots, such as $-0.45<\beta_2<-0.35$, $-0.35<\beta_2<-0.25$, etc.}
\label{fig1}
\end{figure}

In Fig.~\ref{fig1}, the performance of the KRR training process is illustrated with the root-mean-square (rms) deviations $\Delta_{\rm rms}$ of the kinetic and spin-orbit energies between the KRR predictions and the exact values in the validation and test sets.
It can be seen that the rms deviations $\Delta_{\rm rms}$ are generally between 5~keV and 25~keV, depending on the quadrupole deformation $\beta_2$.
This is quite a high accuracy for nuclear physics and would be sufficient for our purpose of studying the shell effects.
More importantly, the rms deviations are quite similar for the validation and test sets, indicating that the KRR network is well trained in its ability to generalize.

Once the orbital-free energy density functional $E_{\rm tot}^{\rm ML}[\rho]$ has been obtained, self-consistent procedures can be performed to find the density that minimizes the total energy using the gradient descent method, starting from a trial density $\rho^{(0)}$.
It should be noted that the center-of-mass (c.m.) correction energies and the Coulomb energies are included in the calculations for comparison with the experimental data, as in Ref.~\cite{Wu2022Phys.Rev.C}.

A random selection of 100 densities from the test sets is employed as the initial density for the self-consistent solution of the ML orbital-free DFT for either $^{16}$O or $^{20}$Ne.
A set of similar ground-state densities and energies is then obtained for each nucleus, and the mean value and standard deviation are taken as the final result and the corresponding statistical uncertainty, respectively.
The self-consistent quadrupole deformations, rms radii, and total energies obtained for $^{16}$O and $^{20}$Ne are compared with the Kohn-Sham results and the available data in Table~\ref{tab1}.

\begin{table}[!htbp]
\caption{Quadrupole deformations $\beta_2$, root-mean-square radii $R_m$ (fm), and total energies $E_{\rm tot}$ (MeV) for $^{16}$O and $^{20}$Ne obtained with the self-consistent machine-learning orbital-free and Kohn-Sham approaches, in comparison with the data available~\cite{NNDC}. }
\begin{center}
\setlength{\tabcolsep}{1.8mm}{
\begin{tabular}{l l r r r}
\hline
\hline
 & & Kohn-Sham & Machine-Learning & Experiment \\
\hline
 $^{16}$O & $\beta_2$       & 0.00      & 0.00 ~  (0.03)    & /         \\
 & $R_m$    & 2.81      & 2.82 ~  (0.02)    & 2.70      \\
 & $E_{\rm tot}$            & -127.45   & -127.40 ~ (0.13)  & -127.62 \\
\hline
 $^{20}$Ne & $\beta_2$      & 0.48      & 0.49    ~ (0.04)  & /       \\
 & $R_m$    & 3.02      & 3.05    ~ (0.02)  & 3.01     \\
 & $E_{\rm tot}$            & -156.58   & -156.02 ~ (0.35)  & -160.64 \\
\hline
\hline
\end{tabular}}
\end{center}
\label{tab1}
\end{table}

The ground-state properties obtained by the present ML approach are in good agreement with the Kohn-Sham ones for both the spherical $^{16}$O and the deformed $^{20}$Ne nuclei.
The deviations for the quadrupole deformations $\beta_2$, rms radii $R_m$, and total energies $E_{\rm tot}$ are small and generally in the range of the statistical uncertainties.
No existing orbital-free DFT method can match this level of performance, particularly when dealing with deformed nuclei.
This is the first successful orbital-free nuclear EDF that has been developed to address the challenge of incorporating deformation shell effects into the framework of orbital-free DFT.

\begin{figure}[!htbp]
\centering
\includegraphics[width=8.0cm]{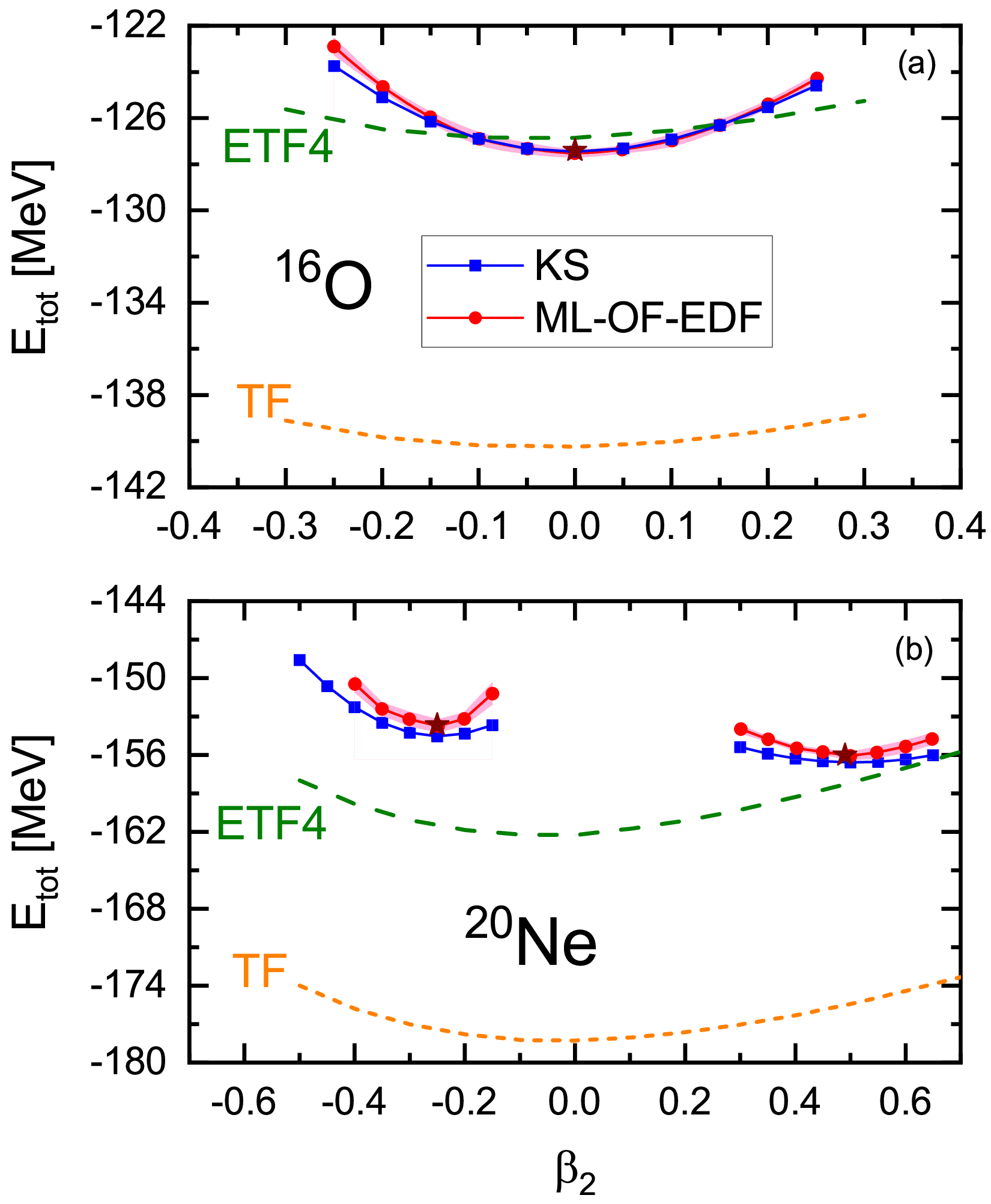}
\caption{Potential energy curves of $^{16}$O (a) and $^{20}$Ne (b) obtained with the machine learning orbital-free DFT, in comparison with the Kohn-Sham and (extended) Thomas-Fermi results.
The stars denote the energy minima obtained by the machine learning orbital-free DFT.
}
\label{fig2}
\end{figure}

Apart from the ground states, self-consistent constrained calculations have also been performed for the potential energy curves of $^{16}$O and $^{20}$Ne.
The results are depicted in Fig.~\ref{fig2}, in comparison with the potential energy curves given by the Kohn-Sham and (extended) Thomas-Fermi approaches.
Prior to the advent of our ML approach, the extended Thomas-Fermi method represented the sole orbital-free DFT available for nuclear systems.
As illustrated in Fig.~\ref{fig2}, the Thomas-Fermi potential energy curves exhibited a notable divergence from the Kohn-Sham curves for both nuclei.
The extended Thomas-Fermi approach (ETF4) has been demonstrated to markedly enhance the results, yet it consistently yields spherical energy minima due to the absence of deformed quantum shell effects. This provides the major challenge of developing the orbital-free density functional theory for nuclear systems.

As illustrated in Fig.~\ref{fig2}, the current ML orbital-free DFT is accurate in reproducing not only the ground-state energies but also the potential energy curves derived from the Kohn-Sham calculations.
In particular, for the deformed $^{20}$Ne, the existence of oblate and prolate energy minima can be reproduced quite well.
It is crucial to highlight that for $^{20}$Ne, as the deformation approaches  $\beta_2=0$, the single-particle levels originating from $1d_{5/2}$ become degenerate. Pairing correlations should be considered in this case in principle. In the absence of pairing correlations, it is challenging to obtain converged results for the potential energy curve in the vicinity of $\beta_2=0$ in Kohn-Sham calculations due to the influence of quantum shell effects. This phenomenon is accurately captured by the present ML orbital-free DFT approach, whereas it cannot be accounted for by Thomas-Fermi approaches.

\begin{figure}[!htbp]
\centering
\includegraphics[width=8.0cm]{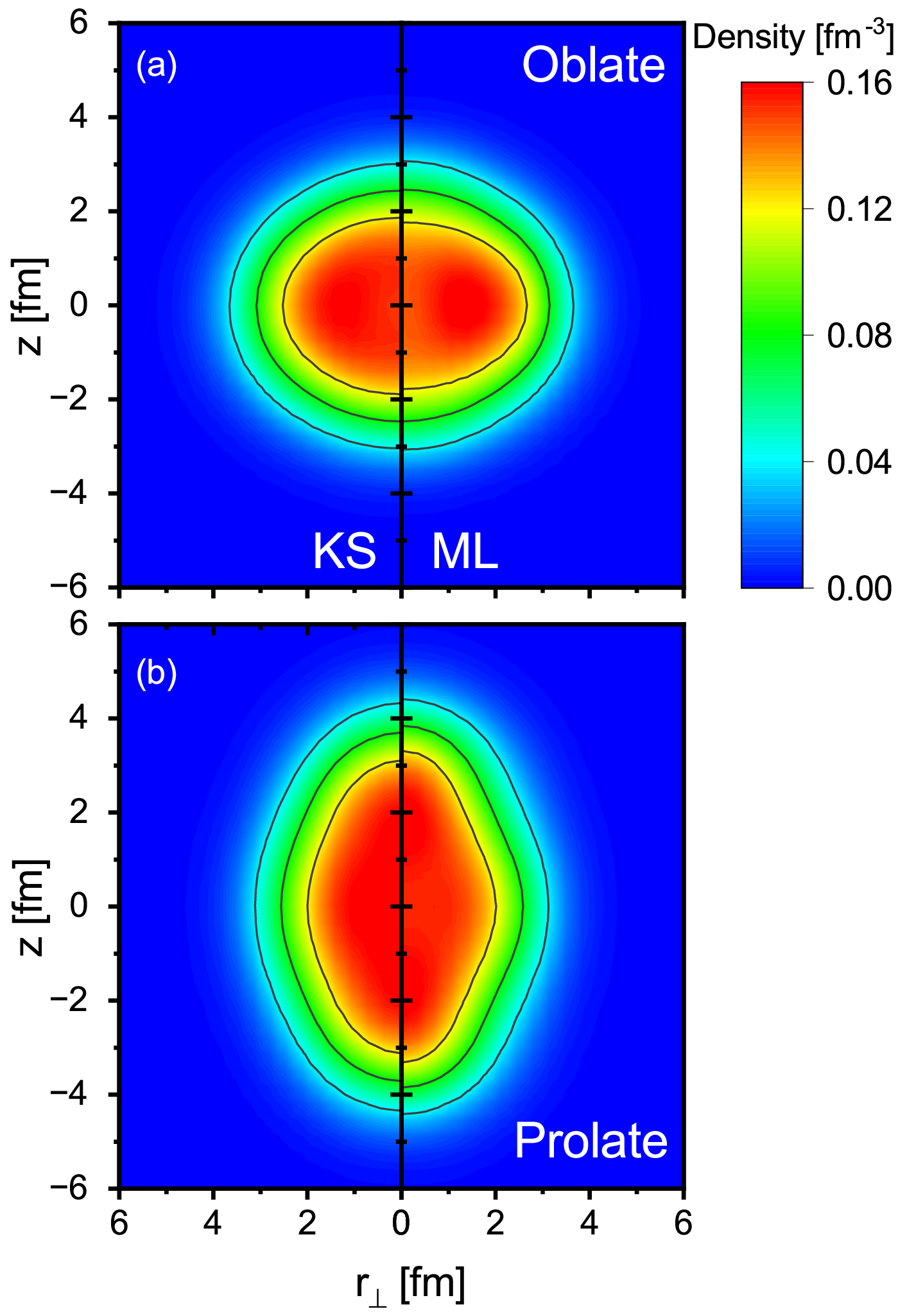}
\caption{Density profiles of the ground state (prolate) and the shape-isomer state (oblate) for $^{20}$Ne.
The results given by the Kohn-Sham DFT and the machine learning orbital-free DFT are shown in the left and right panels, respectively.
}
\label{fig3}
\end{figure}

Finally, the density profiles of the ground state and the shape-isomer state for $^{20}$Ne are depicted in Fig.~\ref{fig3}.
It can be seen that the ML density profiles reproduce the Kohn-Sham densities quite well for both ground and isomeric states.
The spatial fluctuation of the density profiles, corresponding to the quantum shell effects, is well captured by the present ML orbital-free DFT.
All these results demonstrate that the nuclear quantum shell effects are adequately incorporated into the present ML orbital-free DFT.

In summary, the orbital-free density functional theory has been developed for deformed nuclei using a machine-learning approach to construct the kinetic and spin-orbital components.
The constructed machine-learning orbital-free density functional is capable of achieving high accuracy in reproducing the nuclear ground-state properties obtained with the commonly used orbital-dependent Kohn-Sham approach, including deformation, root-mean-square radius, and binding energy.
Furthermore, the current machine-learning orbital-free density functional allows for the accurate calculation of the potential energy curve, thereby facilitating the determination of the isomeric state of nuclei.
The predicted density profiles of the ground state and the isomeric state for $^{20}$Ne reproduce the Kohn-Sham densities with a high degree of precision, and the spatial fluctuations of the density profiles, corresponding to the quantum shell effects, are accurately captured by the present machine-learning orbital-free density functional.
This is the inaugural instance where a fully orbital-free energy density functional has succeeded in taming the complex shell effects in deformed nuclei. It demonstrates that the orbital-free energy density functional, which is directly based on the Hohenberg-Kohn theorem, is not only a theoretical concept but also a practical one.
Therefore, it opens the door to a quantitative study of nuclear systems with the orbital-free density functional theory.



\medskip



{\it This work was partly supported by the National Key R\&D Program of China (Contract No. 2018YFA0404400), the National Natural Science Foundation of China (Grants No. 11935003, No. 11975031, No. 12141501, No. 12070131001, No. 12405134), the start-up Grant No. XRC-23103 of Fuzhou University, the State Key Laboratory of Nuclear Physics and Technology, Peking University (Grant No. NPT2023KFY02), and the China Postdoctoral Science Foundation (Grant No. 2021M700256).
Z.X.R. is supported in part by the European Research Council (ERC) under the European Union's Horizon 2020 research and innovation programme (Grant agreement No. 101018170).
}

\bigskip

\noindent {\bf \large Author Contributions}
Conceptualization, X.H.W. and P.W.Z; methodology, X.H.W. and Z.X.R.; formal analysis, X.H.W.; investigation, X.H.W., Z.X.R. and P.W.Z.; writing—original draft preparation, X.-H.W.; writing—review and editing, All authors; visualization, X.H.W.; supervision, P.W.Z.; funding acquisition, X.H.W. and P.W.Z. 
\\
\\
\noindent {\bf \large Author Information}  The authors declare no competing financial interests.
\\
\\
\noindent {\bf \large Code Availability} All of the codes produced in association with this work have been stored and can be obtained upon request from the authors, subject to possible export control constraints.
\\
\\
\noindent {\bf \large Data Availability} All of the data produced in association with this work have been stored and can be obtained upon request from the authors.
\\
\\
\noindent {\bf \large Inclusion and Ethics} We have complied with community standards for authorship and all relevant recommendations with regard to inclusion and ethics.
\newpage

\section*{Methods}
\renewcommand{\thesection}{S\arabic{section}}
\renewcommand{\thefigure}{S\arabic{figure}}
\renewcommand{\theequation}{S\arabic{equation}}
\renewcommand{\thetable}{S\arabic{table}}
\setcounter{figure}{0}
\setcounter{equation}{0}
\setcounter{section}{0}
\setcounter{table}{0}

\section{Data generation by solving the Schr\"odinger equations}

The data sets of densities, kinetic energies, and spin-orbital energies are built by solving the Schr\"odinger equations with mean potentials $V({\bm r})$ and spin-orbital potentials ${\bm W}({\bm r})$,
\begin{equation}\label{Sch_eq}
  \left[ -\frac{\hbar^2}{2m}\nabla^2 + V + {\bm W}\cdot(-i)(\nabla\times\sigma) \right]\psi = E\psi.
\end{equation}
The mean potential $V$ is simulated by a combination of spherical part $V_0({\bm r})$ and quadrupole-deformed part $V_2({\bm r})$,
\begin{equation}
  V({\bm r}) = V_0(r) + V_2({\bm r}).
\end{equation}
The spherical part is simulated by the combined Gaussian function with 6 parameters,
\begin{eqnarray}\label{potential}
  V_{0}(r) &=&  -a_{01}\exp\left[-\frac{(r-b_{01})^2}{2c_{01}^2}\right] - a_{02}\exp\left[-\frac{(r-b_{02})^2}{2c_{02}^2}\right] \nonumber \\
  &&-\frac{a_{01}(r-b_{01})r}{c_{01}^2}\exp\left[-\frac{(r-b_{01})^2}{2c_{01}^2}\right].
\end{eqnarray}
The combined Gaussian function had been adopted in our previous work~\cite{Wu2022Phys.Rev.C} to simulate the spherical mean potential, and was found to be a flexible smooth function.
The quadrupole-deformed part is simulated by
\begin{equation}
  V_{2}({\bm r}) = \pm V_{20}(r)\cdot P_2(\cos\theta),
\end{equation}
where $P_{2}$ is the second term of Legendre polynomial, and $V_{20}(r)$ is again simulated by the flexible combined Gaussian function,
\begin{eqnarray}\label{potential}
  V_{20}(r) &=&  -a_{21}\exp\left[-\frac{(r-b_{21})^2}{2c_{21}^2}\right] - a_{22}\exp\left[-\frac{(r-b_{22})^2}{2c_{22}^2}\right] \nonumber \\
  &&-\frac{a_{21}(r-b_{21})r}{c_{21}^2}\exp \left[-\frac{(r-b_{21})^2}{2c_{21}^2}\right].
\end{eqnarray}
The corresponding parameters in the mean potential are randomly sampled in the range of $7<a_{01}<9$~MeV, $1.8<b_{01}<2.1$~fm, $0.95<c_{01}<1.15$~fm, $48<a_{02}<59$~MeV, $-0.15<b_{02}<0.25$~fm, $1.7<c_{02}<2.0$~fm,
and $0<a_{21}<6$~MeV, $2.9<b_{21}<3.2$~fm, $0.75<c_{21}<1.05$~fm, $0<a_{22}<12$~MeV, $1.6<b_{22}<1.8$~fm, $0.6<c_{22}<1.1$~fm, respectively.
The spin-orbital potential is given by ${\bm W}({\bm r}) = \frac{3}{4}W_0{\bm \nabla}\rho_{\rm pre}$, where the density $\rho_{\rm pre}$ is pre-calculated by solving the Schr\"odinger equation with ${\bm W}({\bm r}) = 0$.
In addition, the calculated root-mean-square radius $R_{\rm rms}$ of nuclei should be in the range of  $0.8A^{1/3}$ fm and $1.2A^{1/3}$ fm, which is an empirical estimate of nuclear radii according to the basic properties of nuclear force.

In total, 24000 pairs of densities, kinetic energies, and spin-orbital energies for nuclear systems with $^{16}$O and $^{20}$Ne are generated.
These two nuclei are taken as a typical spherical nucleus and a typical deformed nucleus respectively, and they both have the same proton and neutron numbers.
These data are divided into three sets, i.e., 20000 pairs for the training set, 2000 ones for the validation set, and 2000 ones for the test set.
In each set, the numbers of the data for systems with $^{16}$O and $^{20}$Ne are equal.
Since the 24000 pairs of data are generated with mean potentials with quadrupole deformation, the densities have deformations which range in $-0.45 \leq \beta_2\leq 0.65$.

\section{Detail of the functional and its derivative}

The fully orbital-free EDF constructed in this work is expressed as
\begin{equation}
  E_{\rm tot}^{\rm ML}[\rho] = E_{\rm kin+so}^{\rm ML}[\rho] + E_{\rm int}'[\rho].
\end{equation}
For axially deformed systems, it can be written as
\begin{equation}
  E_{\rm tot}[\rho(r_{\perp},z)] = E_{\rm kin+so}^{\rm ML}[\rho(r_{\perp},z)] + E_{\rm int}'[\rho(r_{\perp},z)].
\end{equation}
The interaction part is taken from the Skyrme functional with the form
\begin{align}
  &E_{\rm int}'[\rho(\bm{r})] =   \int\left( \frac{3}{8}t_{0}\rho^2 + \frac{1}{16}t_{3}\rho^{2+\gamma} \right)2\pi r_{\perp}{\rm d}r_{\perp}{\rm d}z \\
  & + \int \frac{1}{64}(9t_1 - 5t_2 -4t_2x_2)\left[\left(\frac{\partial\rho}{\partial r_{\perp}}\right)^2+\left(\frac{\partial \rho}{\partial z}\right)^2\right] 2\pi r_{\perp}{\rm d}r_{\perp}{\rm d}z. \notag
\end{align}
where the parameters are taken from the SkP~\cite{Dobaczewski1984Nucl.Phys.A103}.
The spin-orbital part $E_{\rm so}$ together with the kinetic part $E_{\rm kin}$ are expressed solely as functional of density with the KRR approach,
\begin{equation}
  E_{\rm kin+so}^{\rm ML}[\rho] = \sum_{i=1}^{m} \omega_{i} K(\rho_i,\rho),
\end{equation}
where the density is assumed to be axially symmetric $\rho(r_{\perp},z)$.

The functional derivative is required in the self-consistent calculations with the gradient descent method.
For the interaction part, the functional derivative can be derived as
\begin{align}
  &E_{\rm int}[\rho(\bm{r})] =   \int\left( \frac{3}{8}t_{0}\rho^2 + \frac{1}{16}t_{3}\rho^{2+\gamma} \right)2\pi r_{\perp}{\rm d}r_{\perp}{\rm d}z \notag \\
  & + \int \frac{1}{64}(9t_1 - 5t_2 -4t_2x_2)\left[\left(\frac{\partial\rho}{\partial r_{\perp}}\right)^2+\left(\frac{\partial \rho}{\partial z}\right)^2\right] 2\pi r_{\perp}{\rm d}r_{\perp}{\rm d}z.
\end{align}
For the kinetic and spin-orbit part, it reads
\begin{equation}
  \frac{\delta E^{\rm ML}_{\rm kin}[\rho]}{\delta \rho} =\sum_{i=1}^{m}
  \frac{w_i}{AA_i\sigma^2 }(\rho_i - \rho)K(\rho_i,\rho)\cdot\frac{1}{\Delta V},
\end{equation}
where $\Delta V = \Delta r_{\perp} \Delta z$, which is the volume element in the discrete space.

\section{Self-consistent calculations}

\subsection{Unconstrained calculations}

The nuclear ground state is obtained by a variation of the energy density functional with respect to the density, which is calculated with the gradient descent method starting from a trial density.
In each iteration step $i$, it follows
\begin{equation}
  \rho^{(i+1)} = \rho^{(i)} - \epsilon\left.\frac{\delta E_{\rm tot}[\rho]}{\delta \rho}\right|_{\rho=\rho^{(i)}},
\end{equation}
where $\epsilon$ is a constant to control the step size of iteration, whose value is determined via trading the speed and stability for the convergence.

\subsection{Constrained calculations}

The constrained calculation in the framework of orbital-free DFT is achieved by adding a penalizing energy term to the total energy,
\begin{equation}\label{Etot_cst}
  E'_{\rm tot}[\rho] = E_{\rm tot}[\rho] + E_{\rm cst}[\rho].
\end{equation}
The penalizing energy term is written as
\begin{equation}
  E_{\rm cst}[\rho] = \lambda(q[\rho] - q_0) + c(q[\rho] - q_0)^2,
\end{equation}
referring to the augmented Lagrangian method~\cite{Staszczak2010}.
Where $q$ is the physical quantity, i.e., quadruple moment, that would be constrained, and $q_0$ is the target constrained value of $q$.
$c$ is taken as a constant with proper value, while $\lambda$ is updated in each iteration,
\begin{equation}
  \lambda^{(i+1)} = \lambda^{(i)} + 2c(q^{(i)} - q_0).
\end{equation}

In the calculations with constrained deformation, the physical quantity to be constrained is the quadruple moment,
\begin{equation}
  q[\rho] = \int \rho Q_{20} {\rm d}V,
\end{equation}
where
\begin{equation}
  Q_{20} = \sqrt{\frac{5}{16\pi}}(2z^2-r_{\bot}^2),
\end{equation}
and the quadruple deformation reads
\begin{equation}
  \beta_2 = \frac{4\pi}{3A R_0^2}Q_{20} = \frac{4\pi}{3A R_0^2}\sqrt{\frac{5}{16\pi}}\langle3z^2-r^2\rangle.
\end{equation}
The functional derivative of $q[\rho]$ is
\begin{equation}
  \frac{\delta q[\rho]}{\delta \rho} = Q_{20} = \sqrt{\frac{5}{16\pi}}(2z^2-r_{\bot}^2),
\end{equation}
and thus, the functional derivative of the penalizing energy is
\begin{equation}
  \frac{\delta E_{\rm cst}[\rho]}{\delta \rho} = 2c(q-q_0(\lambda))Q_{20}.
\end{equation}
The state with minimal energy under a certain quadruple deformation $\beta_2$ can thus be obtained by minimizing~\eqref{Etot_cst} with the gradient descent method,
\begin{align}
  \rho^{(i+1)} = & \rho^{(i)} - \epsilon\left.\frac{\delta E'_{\rm tot}[\rho]}{\delta \rho}\right|_{\rho=\rho^{(i)}} \\
  = & \rho^{(i)} - \epsilon\left\{\left.\frac{\delta E_{\rm tot}[\rho]}{\delta \rho}\right|_{\rho=\rho^{(i)}} + \left.\frac{\delta E_{\rm cst}[\rho]}{\delta \rho}\right|_{\rho=\rho^{(i)}} \right\}.
\end{align}

The adaptive functional derivative method, which based on principal components analysis and density renormalization as introduced in Supplementary materials of Ref.~\cite{Wu2022Phys.Rev.C}, is also adopted in the present work to guarantee iteration stability.

\section{Thomas-Fermi approach}

In the Thomas-Fermi (TF) and extended Thomas-Fermi (ETF4) approaches adopted in Fig.2 of the main text, the total kinetic energy is expressed by the functional of the nucleon density,
\begin{subequations}
  \begin{align}
    &E_{{\rm kin},\tau}^{\rm TF}=\frac{1}{2 m_N}\int d^3\bm{r}~\frac{3}{5}(3\pi^2)^{2/3}\rho_\tau^{5/3},\\
    &E_{{\rm kin},\tau}^{\rm ETF4}=\frac{1}{2 m_N}\int d^3\bm{r}~\Bigg{\{}\frac{3}{5}(3\pi^2)^{2/3}\rho_\tau^{5/3}+\frac{1}{36}\frac{(\bm{\nabla}\rho_\tau)^2}{\rho_\tau}+\frac{1}{3}\Delta\rho_\tau\nonumber\\
    &\qquad\qquad +\frac{1}{6480}(3\pi^2)^{-2/3}\rho_\tau^{1/3}\left[8\left(\frac{\bm{\nabla}\rho_\tau}{\rho_\tau}\right)^4-27\left(\frac{\bm{\nabla}\rho_\tau}{\rho_\tau}\right)^2\frac{\Delta\rho_\tau}{\rho_\tau}+24\left(\frac{\Delta\rho_\tau}{\rho_\tau}\right)^2\right] \Bigg{\}},
  \end{align}
\end{subequations}
with $\tau$ represents neutron or proton~\cite{Brack1985Phys.Rep.}.

The potential energy curve of an axially deformed nucleus in the semi-classical TF and ETF4 approaches are obtained based on the restricted density variational method together with the form of Woods-Saxon distribution,
\begin{equation}\label{TFden}
  \rho_{\tau}({\bm r})=\frac{\rho_{0}^{(\tau)}}{1+\exp\left[ \frac{r-\mathcal{R}_{\tau}(\theta)}{a^{(\tau)}} \right]},
\end{equation}
where $\mathcal{R}_{\tau}(\theta)$ in Eq.\eqref{TFden} reads
\begin{equation}
  \mathcal{R}_{\tau}(\theta) = R^{(\tau)}[1+\beta_{20}^{(\tau)}Y_{20}(\theta)].
\end{equation}
The central density $\rho_{0}^{(\tau)}$~is determined from the conservation of particle number, while the surface diffuseness $a^{(\tau)}$, typical radius $R^{(\tau)}$, and deformed parameter $\beta_{20}^{(\tau)}$ are variational parameters to obtain the minimal total energy for each constrained $\beta_2$ in the potential energy curve.

\bibliography{paper}

\end{document}